# Observation of the anomalous Hall effect in a collinear antiferromagnet

Zexin Feng[1,#], Xiaorong Zhou[1,#], Libor Šmejkal[2,3,4,#], Lei Wu[5,6], Zengwei Zhu[5,6], Huixin Guo[1], Rafael González-Hernández[7,2], Xiaoning Wang[1], Han Yan[1], Peixin Qin[1], Xin Zhang[1], Haojiang Wu[1], Hongyu Chen[1], Zhengcai Xia[5,6], Chengbao Jiang[1], Michael Coey[1,8], Jairo Sinova[2,3,*], Tomáš Jungwirth[3,9], Zhiqi Liu[1,*]

1. School of Materials Science and Engineering, Beihang University, Beijing 100191, China.
2. Institut für Physik, Johannes Gutenberg Universität Mainz, D-55099 Mainz, Germany.
3. Institute of Physics, Academy of Sciences of the Czech Republic, Cukrovarnická 10, 162 00 Praha 6 Czech Republic.
4. Faculty of Mathematics and Physics, Charles University in Prague, Ke Karlovu 3, 121 16 Prague 2, Czech Republic.
5. Wuhan National High Magnetic Field Center, Huazhong University of Science and Technology, Wuhan 430074, China.
6. School of Physics, Huazhong University of Science and Technology, Wuhan 430074, China.
7. Grupo de Investigacíon en Física Aplicada, Departamento de Física, Universidad del Norte, Barranquilla, Colombia.
8. Department of Pure and Applied Physics, Trinity College, Dublin 2, Ireland.
9. School of Physics and Astronomy, University of Nottingham, Nottingham NG7 2RD, United Kingdom.

[#]These authors contributed equally to this work.

*emails: zhiqi@buaa.edu.cn (experiment); sinova@uni-mainz.de (theory)

**Time-reversal symmetry breaking is the basic physics concept underpinning many magnetic topological phenomena such as the anomalous Hall effect (AHE) and its quantized variant[1-5]. The AHE has been primarily accompanied by a ferromagnetic dipole moment, which hinders the topological quantum states and limits data density in memory devices[1,6-8], or by a delicate noncollinear magnetic order[9-14] with strong spin decoherence, both limiting their applicability. A potential breakthrough is the recent theoretical prediction of the AHE arising from collinear antiferromagnetism in an anisotropic crystal environment[15]. This new mechanism does not require magnetic dipolar or noncollinear fields. However, it has not been experimentally observed to date. Here we demonstrate this unconventional mechanism by measuring the AHE in an epilayer of a rutile collinear antiferromagnet $RuO_2$. The observed anomalous Hall conductivity is large, exceeding 300 S·cm$^{-1}$, and is in agreement with the Berry phase topological transport contribution[15-17]. Our results open a new unexplored chapter of time-reversal symmetry breaking phenomena in the abundant class of collinear antiferromagnetic materials.**

Anomalous Hall effect (AHE) is a text-book time-reversal symmetry breaking (TRSB) phenomenon, which remains at the forefront of the research of topological magnetism and dissipationless nanoelectronics[1,3,4]. The AHE current is transverse to the applied electric field and is odd under time-reversal. The phenomenon is microscopically understood as a consequence of a spin-splitting of electronic states by the magnetic ordering, which is linked to the electrons' momenta by the spin-orbit coupling[1,18]. An intrinsic, scattering-independent contribution to AHE is of topological origin, given by the momentum-space Berry curvature of the electron Bloch states[1,16,17].

The bulk of the AHE literature is focused on TRSB by a ferromagnetic order, illustrated in Fig. 1a by the Fe crystal[1,7,8]. In collinear ferromagnets like Fe, a majority of electrons occupy one spin state, resulting in a macroscopic dipole moment aligned with the majority spin axis. In the past two decades, AHE was demonstrated also as a consequence of TRSB by a noncollinear magnetic order. The studies included ferromagnets[9,12], as well as antiferromagnets[11,13,14,19-22]. For example, in the $Mn_3X$ family of noncollinear antiferromagnets (X = Ir, Sn, Pt)[13,14,20], illustrated in Fig. 1b, the magnetic dipole is weak, while strong TRSB effects and AHE were associated with the presence of a magnetic octupole[23]. However, both dipolar and noncollinear magnetic orders represent an obstacle for realizing quantized variants of the anomalous Hall systems, and for their practical utility in nanoelectronic devices. On one hand, the dipolar fields disturb the topological quantum states, or limit the data density in magnetic memories. The limitations imposed by the magnetic dipole can be remedied in the (nearly) compensated noncollinear antiferromagnets. On the other hand, this comes at the cost of strongly decohered spin quantum numbers of the electrons. Moreover, the noncollinear states are delicate due to frustrated magnetic interactions in the lattice, the presence of competing magnetic phases, and the sensitivity to variations in structure parameters, such as the film thickness[24] or disorder[11,13].

To date, an anomalous Hall current in collinear antiferromagnets has been experimentally identified only as a consequence of canting of the magnetic moments by an applied magnetic field, or due to a field-induced spin-flip transition into a ferromagnetic state[6,25,26]. In both cases, AHE was ascribed to the magnetic dipole in the material, generated by the applied magnetic field[26]. In the absence of an external magnetic field, collinear antiferromagnets were commonly perceived to exclude AHE. This is indeed the case when the opposite magnetic sublattices are connected by time-reversal combined with a spatial symmetry operation (*e.g.*, inversion or half-unit cell translation)[11,19,27,28]. The resulting effective time-reversal symmetry, $T_{AF}$, prohibits the AHE.

Recently, this general perception has been broken by Smejkal *et al.*[15], who predicted large AHE arising from a previously overlooked TRSB mechanism, generated by a collinear antiferromagnetic order in an anisotropic crystal environment. The mechanism applies to a broad materials class of collinear antiferromagnetic crystals allowing, by symmetry, also for a ferromagnetic dipole. Since the dipole moment is weak, earlier studies assumed that the response of these so called weak ferromagnets was close to that of the perfectly compensated $T_{AF}$-symmetric antiferromagnets[29]. However, taking collinear antiferromagnet $RuO_2$ as an example, Smejkal e*t al.*[15] have theoretically demonstrated that the weak ferromagnetism and AHE are of a fundamentally distinct microscopic origin. The weak ferromagnetism is due to relativistic spin-orbit fields and represents only a perturbative correction[29,30]. In contrast, there is an additional strong TRSB mechanism arising from the collinear magnetic-exchange fields and the anisotropic crystal structure, which produces large AHE even when the ferromagnetic dipole moment vanishes[15,31,32].

Before turning to our experimental observation of large AHE in the $RuO_2$ antiferromagnet, we present density functional theory (DFT) calculations addressing the relevant experimentally geometry. Specifically, we show that for the Néel vector aligned with the [110] crystal axis, the collinear antiferromagnet has a large Berry curvature and corresponding intrinsic AHE. In the first step, we demonstrate strong unconventional TRSB on non-relativistic spin-split bands of $RuO_2$[15,33]. In the second step we show how this TRSB is translated into the momentum-dependent Berry curvature when turning on the relativistic spin-orbit coupling.

DFT calculations in Fig. 1c illustrate that TRSB seen in the local magnetization densities of $RuO_2$ corresponds to the symmetry lowering of magnetic Ru sites by the presence of the non-magnetic oxygen atoms. Here, the $T_{AF}$-symmetry connecting the antiferromagnetic sublattices is broken by the oxygen octahedra and generates the low-symmetry magnetization isosurfaces, shown in Fig. 1c. These anisotropic real-space magnetization densities, dominated by $d_{xz}$ and $d_{yz}$ orbitals of Ru, are directly linked in the momentum-space to a strong spin-splitting of non-relativistic energy bands, reaching 1 eV around the Fermi level, as shown in Fig. 1d,f and Supplementary Fig. 1 (ref. [15,33-35]).

The oxygen octahedra formed around the two Ru atoms in the unit cell are rotated with respect to each other by 90° and this symmetry is related to the perfect spin-compensation, *i.e.*, the lack of a magnetic dipole in the (non-relativistic) spin-split bands. Instead of a dipole, the RuO$_2$ spin-split Fermi surfaces[15], shown in Fig. 1d, have the characteristic form of a magnetic toroidal quadrupole[36]. This is in contrast to a conventional dipolar spin-splitting, which is illustrated in Fig.1e on a hypothetical ferromagnetic phase of RuO$_2$. In addition, the spin-component along the Néel vector is a good quantum number in the RuO$_2$ band structure, *i.e.*, commutes with the Hamiltonian, in the absence of the relativistic spin-orbit coupling. This is highlighted in Fig.1d,f where the bands are labeled by the up and down-spin quantum numbers. The collinear antiferromagnetic TRSB in RuO$_2$, therefore, shares the vanishing dipole magnetic moment with the non-collinear antiferromagnetic TRSB, while keeping the favorable spin-conserving nature of ferromagnetic TRSB.

When we connect the spin and momentum sectors by adding spin-orbit coupling in our DFT calculations, the RuO$_2$ antiferromagnet can exhibit an anomalous Hall vector $\boldsymbol{\sigma}_{\text{AHE}}$: $\boldsymbol{j}_{\text{AHE}} = \boldsymbol{\sigma}_{\text{AHE}} \times \boldsymbol{E}$, where $\boldsymbol{j}_{\text{AHE}}$ is the anomalous Hall current, $\boldsymbol{E}$ is the applied voltage, and $\boldsymbol{\sigma}_{\text{AHE}} = (\sigma_{yz}, \sigma_{zx}, \sigma_{xy})$. The presence and orientation of $\boldsymbol{\sigma}_{\text{AHE}}$ is sensitive to the orientation of the antiferromagnetic Néel vector $\boldsymbol{N} = (\boldsymbol{M}_A - \boldsymbol{M}_B)/2$ (ref. [15]). In Fig. 1g we show our DFT calculation of the crystal momentum-resolved Berry curvature[15] for the antiferromagnetic Néel vector oriented along the [110] direction. We observe hotspots where the Berry curvature is larger in magnitude than in ferromagnetic Fe[26], or in noncollinear antiferromagnets[14]. By integrating the Berry curvature, we obtain the intrinsic anomalous Hall conductivity which shows large values up to 1000 S/cm, depending on the position of the Fermi level (see Supplementary Fig. 2,3). This indicates the possibility of unexpectedly large Berry curvature effects, and particularly anomalous Hall conductivity, in RuO$_2$ with the Néel vector aligned along the [110] crystal direction[15].

Rutile RuO$_2$ is a good metal with a room-temperature resistivity of only $\rho \sim 35$ μΩ·cm (ref. [37]). Recent neutron scattering[38] and resonant X-ray scattering[39] studies demonstrated the presence of collinear antiferromagnetism in both its thin films and bulk crystals above room temperature, where the spin axis is slightly tilted off the *c*-axis. It has been previously shown[40] that thin-film growth of rutile RuO$_2$ on oxide single-crystal MgO and SrTiO$_3$ substrates by pulsed laser deposition yields highly-quality (110)-oriented and (100)-oriented epitaxial films, respectively, which facilitates the investigation of the AHE in $RuO_2$ thin films along different crystallographic orientations.

To explore the AHE experimentally, we fabricated epitaxial thin films of $RuO_2$. To optimize the electrical conductivity, systematic growth of $RuO_2$ onto MgO single-crystal substrates was performed for a wide range of oxygen partial pressure between $10^{-2}$ and $10^{-6}$ Torr and a wide growth temperature range between 500 and 800°C. The thickness of the $RuO_2$ films was kept at 27 nm. The optimal growth conditions are 550°C and $10^{-3}$ Torr oxygen pressure, for which the lowest room-temperature resistivity of $\rho \sim 64$ μΩ·cm was achieved (Supplementary Fig. 4). Although it is higher than the bulk resistivity, the value for rutile $RuO_2$ films is one fifth of the resistivity of a commonly used perovskite metallic oxide $SrRuO_3$ (ref. [41]). Generally, $RuO_2$ thin films grown below 700°C are all epitaxial and exhibit metallic transport behavior (Supplementary Fig. 4).

Figure 2a shows an X-ray diffraction pattern of an optimized 27-nm-thick $RuO_2$/MgO heterostructure and reveals the (110) orientation of epitaxial $RuO_2$, which is consistent with the previous growth[40]. The quality of our films is also evidenced by the cross-section transmission electron microscopy image in Fig. 2b. Magnetization measurements indicate a weak net moment for such $RuO_2$ thin films reaching 0.2 $\mu_B$ per unit cell at 50 T, as shown in Fig. 2c (see Methods).

To demonstrate the presence of the antiferromagnetic order in our $RuO_2$ films, we performed magnetization measurements of a $Co_{90}Fe_{10}$/$RuO_2$ exchange bias stack. The soft magnetic layer of $Co_{90}Fe_{10}$ with a thickness of 5 nm and a 2-nm-thick Pt cap to prevent $Co_{90}Fe_{10}$ from oxidation were deposited on top of the $RuO_2$ (Fig. 2d-f) at room temperature in a magnetic field of ~20 mT, yielding a clear exchange bias at 50 K (Fig. 2f). This agrees with the reported antiferromagnetic order of rutile $RuO_2$ films revealed by resonant X-ray scattering studies[39]. The interfacial exchange coupling between $RuO_2$ and $Co_{90}Fe_{10}$ causes a large enhancement of the coercivity field of $Co_{90}Fe_{10}$ (Fig. 2d). The blocking temperature of ~200 K (Fig. 2f) could be increased towards the Néel temperature of $RuO_2$ by post-annealing in a magnetic field[42].

As shown below, the AHE is suppressed in otherwise similar $RuO_2$ films deposited onto $SrTiO_3$. These films, therefore, serve as crucial reference systems in our experimental AHE study. The $RuO_2$ films deposited onto $SrTiO_3$ at the same optimized conditions as for the $RuO_2$/MgO growth are highly ordered as well, but (100)-oriented (Supplementary Fig. 5). Magnetization measurements display again a weak net moment reaching ~0.2 $\mu_B$ per unit cell at 50 T (Supplementary Fig. 6).

In Fig. 3 we compare transport measurements on (110)-oriented $RuO_2$/MgO and on (100)-oriented $RuO_2$/$SrTiO_3$. The temperature-dependent longitudinal resistivity is nearly identical in the two film-orientations (Fig. 3a). The two orientations have also the same Néel transition temperature which is close to 350 K, as inferred from the transport anomaly[43] highlighted in the inset of Fig. 3a by the peak in d$\rho$/d$T$. (We note that the Néel temperature above 300 K in our films is consistent with earlier reports[38,39]). Figure 3b shows the longitudinal magnetoresistance measured with out-of-plane fields up to 50 T. The observed form corresponds to the ordinary positive magnetoresistance due to the Lorentz force, typical in metals. Again, the plots are nearly identical in the $RuO_2$/MgO and $RuO_2$/$SrTiO_3$ films over the entire range of applied fields.

In in the inset of Fig. 3c, we plot Hall measurements in a magnetic field applied along the out-of-plane direction of the thin films at 340 K. As for the longitudinal magnetoresistance, these high-temperature linear Hall signals in the two film-orientations are identical and can be ascribed to the ordinary Lorentz force contribution. However, at lower temperatures, a clear anomaly appears in the Hall signal of (110)-oriented $RuO_2$/MgO, while the Hall data remain linear in (100)-oriented $RuO_2$/$SrTiO_3$, as shown for 10 K in the main plot of Fig. 3c. This is highlighted in Fig. 3d by subtracting the linear fit to the total Hall data between -40 and -50 T. The nonlinear Hall contribution is seen in the (110)-oriented $RuO_2$/MgO film (while it is absent in $RuO_2$/$SrTiO_3$) over an entire range of temperatures from 10 to 300 K (Supplementary Fig. 7 & 8).

We show below that the nonlinear Hall signal is consistent with TRSB and the AHE described in theoretical Fig. 1. Specifically, we present the following set of arguments: (i) The AHE signal has the theoretically expected magnitude, sign, and temperature dependence. (ii) The (110)-oriented $RuO_2$/MgO and (100)-oriented $RuO_2$/$SrTiO_3$ films have identical ordinary longitudinal and Hall signals which confirms that the additional contribution seen only in the (110)-oriented film is of the anomalous origin, *i.e.*, unrelated to the ordinary Lorentz force Hall effect. (iii) The observed anisotropy of AHE is again consistent with the theoretical phenomenology. This includes the vanishing anomalous Hall signal for the Néel vector aligned with the [001] easy-axis and the corresponding absence of a finite and hysteretic anomalous Hall signal at zero magnetic field. (iv)

Related to the previous point, the strong magnetic field required for observing AHE is only needed for reorienting the Néel vector away from the easy-axis. It is not required, and up to the measured 50 T it is not contributing significantly, to the TRSB mechanism of AHE. In addition, also the spin-flip transition is excluded since the hypothetical ferromagnetic state (Fig.1b) has a much higher energy than the antiferromagnetic ground state (by 56 meV per unit cell), and has a net magnetization of ~0.4 $\mu_B$ per unit cell which we do not observe in our experiment in Fig. 2c. We now discuss these points in detail.

Considering the extracted anomalous Hall resistivity $\rho_{AHE}$ and the longitudinal (zero-field) resistivity $\rho$ (Fig. 3a) of the RuO$_2$/MgO film at different temperatures, the anomalous Hall conductivity estimated from $\sigma_{AHE} \approx -\rho_{AHE}/\rho^2$ is plotted as a function of temperature in Fig. 3e. At low temperatures, $\sigma_{AHE}$ reaches ~330 S·cm$^{-1}$. This is over three times that of the non-collinear antiferromagnet Mn$_3$Sn[13] (see Supplementary Tab. 1 for a detailed comparison) and even on the same order as the anomalous Hall conductivity of Fe thin films[44]. Above 50 K, $\sigma_{AHE}$ falls off rapidly with increasing temperature, to approximately 3 S·cm$^{-1}$ at room temperature. We comment on the temperature dependence below in the theory discussion.

The nonlinear Hall signal in RuO$_2$/MgO could in principle occur due to the ordinary contribution from the orbital Lorentz force combined with a multi-band transport, or multiple types of carriers or anisotropic Fermi surfaces. However, the measured ordinary longitudinal magnetoresistance shows no significant anisotropy between RuO$_2$/MgO and RuO$_2$/SrTiO$_3$ films (Fig. 3b). Hence, the ordinary Lorentz force contributions to transport are the same in the two film orientations, implying that the observed anomalous nonlinear Hall contribution in RuO$_2$/MgO cannot be ascribed to the ordinary Hall effect. Consistent with this picture, the total measured Hall slopes for the two RuO$_2$ film orientations approach each other at 340 K where the anomalous Hall contribution in (110)-oriented RuO$_2$/MgO approaches zero. We can therefore ascribe the high-temperature linear Hall data above 340 K in both films to the ordinary Hall effect. Since the ordinary longitudinal magnetoresistance shows no anisotropy between the two film orientations down to 10 K, we can also expect that their ordinary Hall contributions are similar down to low temperatures. The corresponding carrier densities at different temperatures, assuming a single-band approximation, are given in Supplementary Tab. 2.

By subtracting the 10 K linear Hall data measured in (100)-oriented RuO$_2$/SrTiO$_3$ from the total 10 K Hall signal in (110)-oriented RuO$_2$/MgO, we obtain an anomalous Hall conductivity estimate which is about a factor of 2 larger than the 10 K value plotted in Fig. 3e. The comparison of the two estimates suggests that the 10 K anomalous Hall component is not fully saturated at 50 T.

Finally, we found that (110)-oriented RuO$_2$ films can be also synthesized by using (011)-oriented PMN-PT (0.7PbMg$_{1/3}$Nb$_{2/3}$O$_3$–0.3PbTiO$_3$) single-crystal substrates. However, these films are less ordered as revealed by X-ray diffraction and transmission electron microscopy characterizations (Supplementary Fig. 9). Although a deviation from the linear ordinary Hall effect can be clearly seen at low temperatures (Supplementary Fig. 10), the effect is much smaller than that of highly-ordered RuO$_2$/MgO films, as shown in Fig. 3e. This suggests that the crystallinity is a key and therefore an effective tuning factor in determining AHE in this collinear antiferromagnet.

We show next, by systematic DFT calculations (see Methods), that our experimental observations are consistent with the theoretically predicted AHE in RuO$_2$ (ref. [15]). Our calculations indicate that the experimentally observed anomalous Hall signal originates from the collinear antiferromagnetism[15] and the external magnetic field is required only for the reorientation of the magnetic moments to the directions with a large anomalous Hall conductivity, *i.e.*, off the [001] easy-axis. Note that an analogous magnetic reorientation was previously shown in a similar collinear antiferromagnet CoF$_2$[45]. The small net magnetic moment represents only a minor correction to the anomalous Hall conductivity and the anomalous Hall conductivity is primarily independent of this weak magnetization, as shown earlier by Smejkal *et al.*[15]. In Fig. 4a we show DFT energy bands for the Néel vector along the [110] axis, which were used for the Berry curvature and anomalous Hall conductivity calculations. We have further verified by DFT calculations (see Methods) that the large experimental magnetic fields have only a small effect on these energy bands, as we shown in Supplementary Fig. 2. The magnetic fields have also a small effect on the net magnetization. For the Néel vector along the [110] axis and Hubbard $U$ = 1.6 eV, we obtain a net magnetization of 0.08 $\mu_B$ at zero field, and 0.09 $\mu_B$ at 35 T. (We note that the majority of the net moment (~80%) comes from the orbital part.) Furthermore, we show in Fig. 4b the dependence of the net moment and Néel vector (sublattice moment) on Hubbard $U$. We see that the values of the total net moment ~0.1 $\mu_B$/Ru correspond well to the experiment value ~0.2 $\mu_B$ per unit cell shown in Fig. 2c.

In Fig. 4c we plot the dependence of the calculated anomalous Hall conductivity on the Fermi level, where zero corresponds to a neutrality point in the employed DFT scheme (see Methods). When shifting the Fermi level to ~0.1 eV, we obtain $\sigma_{AHE}^{DFT} \sim 300$ S/cm, which corresponds in both the magnitude and sign to the experiment at low temperatures. The small Fermi level shift to match the experiment can be ascribed to disorder effects but is also within a numerical scatter seen between different DFT implementations. The position of energy bands around the Fermi level depends on the correlation strength and approximations used in the DFT calculations. We evaluate the Hall conductivity from a Wannier Hamiltonian constructed from pseudopotential calculations with Hubbard $U$ included via the Dudarev scheme (see Methods). However, the Fermi level can shift by ~0.1 eV towards the bands corresponding to the peak in AHE when we use the full potential calculation within the fully localized limit, as we show in Supplementary Fig. 2. In Supplementary Fig. 3 we show the corresponding out-of-plane and in-plane magnetic anisotropy energies (MAE). We see that the large values of the anomalous Hall conductivity correspond to an out-of-plane MAE $K_z \approx 0.1 - 5$ meV. These values are consistent with the antiferromagnetic Néel vector along the [001] axis and a vanishing anomalous Hall signal[15] when there is no applied magnetic field.

Figure 4d demonstrates that to achieve a large anomalous Hall conductivity it is sufficient to tilt the Néel vector out of the [001] direction (not necessarily to fully reorient it into the (001)-plane). In a simple uniaxial antiferromagnet with zero net moment in the ground state, the typical scale of the reorienting magnetic field is given by the spin-flop field, $\sqrt{2K_zJ}$, where $K_z$ and $J$ are the anisotropy and exchange fields, respectively. The exchange field can be estimated from the Néel temperature $J \sim \frac{3}{2} k_B T_N \sim 350$ T for $T_N \sim 350$ K (see Fig. 3a). The calculated anisotropy energies shown in Supplementary Fig. 3 are on the order of ~0.25 meV ($K_z \sim 4$ T), giving $\sqrt{2K_zJ} \sim 50$ T. In the experiment, we see a sizable AHE signal already below 50 T for the (110)-oriented $RuO_2$/MgO film. We ascribe this to the weak equilibrium moment of $RuO_2$ which from the calculations is parallel to the Néel vector when the Néel vector is reoriented along the [110] axis, in effect having a ferrimagnetic order. The presence of the weak moment and the corresponding additional Zeeman energy gain can facilitate the Néel vector reorientation by fields along the [110] axis of magnitudes below the spin-flop scale, similarly as in $CoF_2$[45]. We also point out that the weak net moment in $RuO_2$ allows for selecting one of the two domains with opposite Néel vector along the [110] axis. For a given sign of the magnetic field applied along the [110] axis, one domain will have a lower energy while the opposite domain will prevail when reversing the sign of the magnetic field[15].

While the measured absence of AHE in (100)-oriented $RuO_2$/$SrTiO_3$ allowed us to use this film as a reference when experimentally assigning the non-linear Hall signal in the (110)-oriented $RuO_2$/MgO film to the AHE, we have not identified the primary reason for the suppressed AHE in $RuO_2$/$SrTiO_3$. The differences between $RuO_2$/$SrTiO_3$ and $RuO_2$/MgO films, that could potentially explain the unequal AHE response, include the in-plane MAE (see Supplementary Fig. 3) and the corresponding anisotropy in the magnetic-field reorientation behavior between the two films, anisotropy of AHE itself (see Supplementary Fig. 3), or the presence of opposite crystal-chirality grains leading to a cancellations in the net AHE signal.

The steep drop of the anomalous Hall conductivity with increased temperature in $RuO_2$/MgO can be explained by the decreased magnitude of the Néel vector. In Fig. 4b,e we demonstrate the correlation between the anomalous Hall conductivity and the magnitude of the Néel vector which we control by tuning Hubbard $U$ for the antiferromagnetic vector along the [110] axis. We observe that for small and large Hubbard $U$, the Hall conductivities are negligible. This is due to the vanishing antiferromagnetic moment for small $U$ and opening of an insulating bandgap for large $U$, respectively. Values of $U \approx$ 1.6-2.0 eV correspond to the antiferromagnetic Néel vector magnitude consistent with previous DFT studies[15,33,38,39]. Decreasing $U$ simulates the quenching of the antiferromagnetic moment with increasing temperature. Figure 4e shows that the anomalous Hall conductivity steeply decreases with decreasing $U$, which is consistent with the experimentally observed drop in the Hall conductivity at high temperatures.

In conclusion, by combining epitaxial growth of $RuO_2$ thin films, pulsed high-field Hall measurements and DFT calculations, we observe the AHE associated with collinear antiferromagnetic TRSB. The experimental identification of the AHE in a collinear antiferromagnet removes the, often limiting, requirements of a net moment or a complex magnetic structure for observing this prominent topological phenomenon. Simultaneously, $RuO_2$ is only one example of a broad family of materials allowing for the collinear antiferromagnetic TRSB mechanism and strong spin splitting[15,33-35]. Our work thus opens new directions in frontier materials research of magnetic topological insulators, axion insulators, or quantum AHE systems in the abundant class of collinear antiferromagnets, many of which have high ordering temperatures and are composed of common elements.

## Materials and Methods

**Growth:** $RuO_2$ thin films were first grown onto (001)-oriented MgO single-crystal substrates at different growth oxygen pressures ranging from $10^{-2}$ to $10^{-6}$ Torr and different growth temperatures between 500 and 800°C by pulsed laser deposition with a based pressure of $1.5\times10^{-8}$ Torr. The target-substrate distance was 60 mm. The laser fluence was ~1.6 J/cm$^2$ and the repetition rate was kept at 10 Hz during deposition. The ramp rate was 20 °C/min for heating and 10 °C/min for cooling. The growth of $RuO_2$ thin films onto (001)-oriented $SrTiO_3$ and (011)-oriented $0.7PbMg_{1/3}Nb_{2/3}O_3$–$0.3PbTiO_3$ (PMN-PT) single-crystal substrates was performed at 550 °C and $10^{-3}$ Torr with the same laser fluence of ~1.6 J/cm$^2$ and repetition rate of 10 Hz. Room-temperature growth of the ferromagnetic $Co_{90}Fe_{10}$ and the capping Pt thin films was carried out by a d.c. sputtering system with a base pressure of $7.5\times10^{-9}$ Torr. For the $Co_{90}Fe_{10}$ deposition, the d.c. sputtering power was 90 W and the Ar pressure was 3 mTorr. The growth rate was ~0.11 Å/s. For Pt sputtering, the power was by 30 W and the Ar pressure was 3 mTorr. The growth rate was determined to be 0.5 Å/s. No chemical treatment was taken for substrates before deposition.

**X-ray diffraction:** X-ray spectra were collected in an XRD-SmartLab diffractometer. The Cu-$Ka$ X-ray wavelength 1.541882 Å.

**Transmission electron microscopy:** The focused ion beam technique was used to fabricate cross-section samples. Afterwards, the transmission electron microscopy characterization was conducted in a FEI Talos F200X setup under 200 kV.

**Electrical measurements:** Electrical contacts were fabricated by wire bonding through Al wires with a diameter of 25 μm. The linear four-probe geometry with a space of 1 mm was utilized for resistivity measurements of $RuO_2$ thin films, which were carried out in a Quantum Design physical property measurement system with a measuring current of 1 mA. The typical two-probe resistance for two 1-mm-far bonds was ~30 Ω and the absolute four-probe resistance was ~10 Ω for optimized $RuO_2$ thin films.

**Pulsed high-field Hall measurements:** The conventional Hall geometry was established by Cu wires at Wuhan National High Magnetic Field Center, Huazhong University of Science and Technology, Wuhan, China. The magnetic field was applied along the out-of-plane direction of thin film samples. The amplitude of the a.c. current was 3 mA and the frequency was 100 kHz. The Hall voltage collection was performed by a National Instruments PXIe 5105 oscilloscope at a sampling frequency of 4 MHz.

**Pulsed high-field magnetic moment measurements:** Pulsed-field magnetic moment signals were collected by a pickup coil coaxial with the pulsed magnetic field and calibrated by low-field magnetic moment obtained by a Quantum Design superconducting quantum interference device. To obtain the high-field magnetic moment of the $RuO_2$ thin film, both $RuO_2$/MgO ($SrTiO_3$) heterostructures and MgO ($SrTiO_3$) substrates of the same dimensions were measured, and the difference in magnetic moments of the heterostructure and the substrate was subsequently extracted to represent the thin film signal.

**Density functional theory calculations:** We performed the density functional theory (DFT) calculations employing the projector augmented plane wave method[46] implemented in VASP code and we used the spherically symmetric Dudarev DFT+$U$ (ref. [47]). We set the energy cut-off of the plane-wave basis to 520 eV, the PBE exchange-correlation functional[48], and the crystal momentum grid 16×16×24. We used DFT relaxed lattice parameters $a = b = 4.5337$ Å, $c$ = 3.124 Å and we set the antiferromagnetic moments along the [110] axis. We constructed the maximally localized Wannier functions in the Wannier90 code[49] and we calculated the intrinsic Hall conductivity by employing the Berry curvature formula[15-17]. We used the fine-mesh of 321×321×321 Brillouin zone sampling points. The influence of the magnetic field via the Zeeman effect and electronic correlation method on the electronic structure was studied by calculations in full potential ELK[50] code with the same parameters as in VASP.

**Data availability:** The data and simulations codes that support the findings of this study are available from the corresponding author upon request.

## Acknowledgements

Z.L. acknowledges financial support from the National Natural Science Foundation of China (NSFC; grant numbers 51822101, 51771009 & 11704018). Z.L. & Z.Z. acknowledge financial support from the National Natural Science Foundation of China (NSFC) on the Science Foundation Ireland (SFI)-NSFC Partnership Programme (NSFC Grant No. 51861135104). M.C. acknowledges support from Science Foundation Ireland contract 12/RC/2278. L.S., J.S. & T.J. were supported in part by the Ministry of Education of the Czech Republic Grants LM2018110 and LNSM-LNSpin, by the Grant Agency of the Czech Republic Grant No. 19-28375X, by the EU FET Open RIA Grant No. 766566. LS and RHG acknowledge the use of the supercomputer Mogon at JGU (hpc.uni-mainz.de), the computing and storage facilities owned by parties and projects contributing to the National Grid Infrastructure MetaCentrum provided under the programme "Projects of Large Research, Development, and Innovations Infrastructures" (CESNET LM2015042). LS and JS acknowledge the support from the Alexander von Humboldt Foundation, and the Deutsche Forschungsgemeinschaft (DFG, German Research Foundation) - TRR 173 – 268565370 (project A03).

## Author contributions

Z.F. & Xiaorong Z. performed sample growth, electrical, structural and magnetic measurements with assistance from L.W., Z.Z., H.G., X.W., H.Y., P.Q., Xin.Z., H.W., H.C., Z.X. & C.J. Theoretical calculations and analysis were performed by L.S., R.G.H., J.S. & T.J. The manuscript was written by Z.L., Z.F. Xiaorong Z., M.C., L.S., J.S. & T.J. All authors commented on manuscript. This project was conceived and led by Z.L.

## Competing interests

The authors declare no competing financial interests.

## Additional information

**Correspondence and requests for materials** should be addressed to Z.L. (email: zhiqi@buaa.edu.cn) and J.S. (email: sinova@uni-mainz.de).

# Figure 1

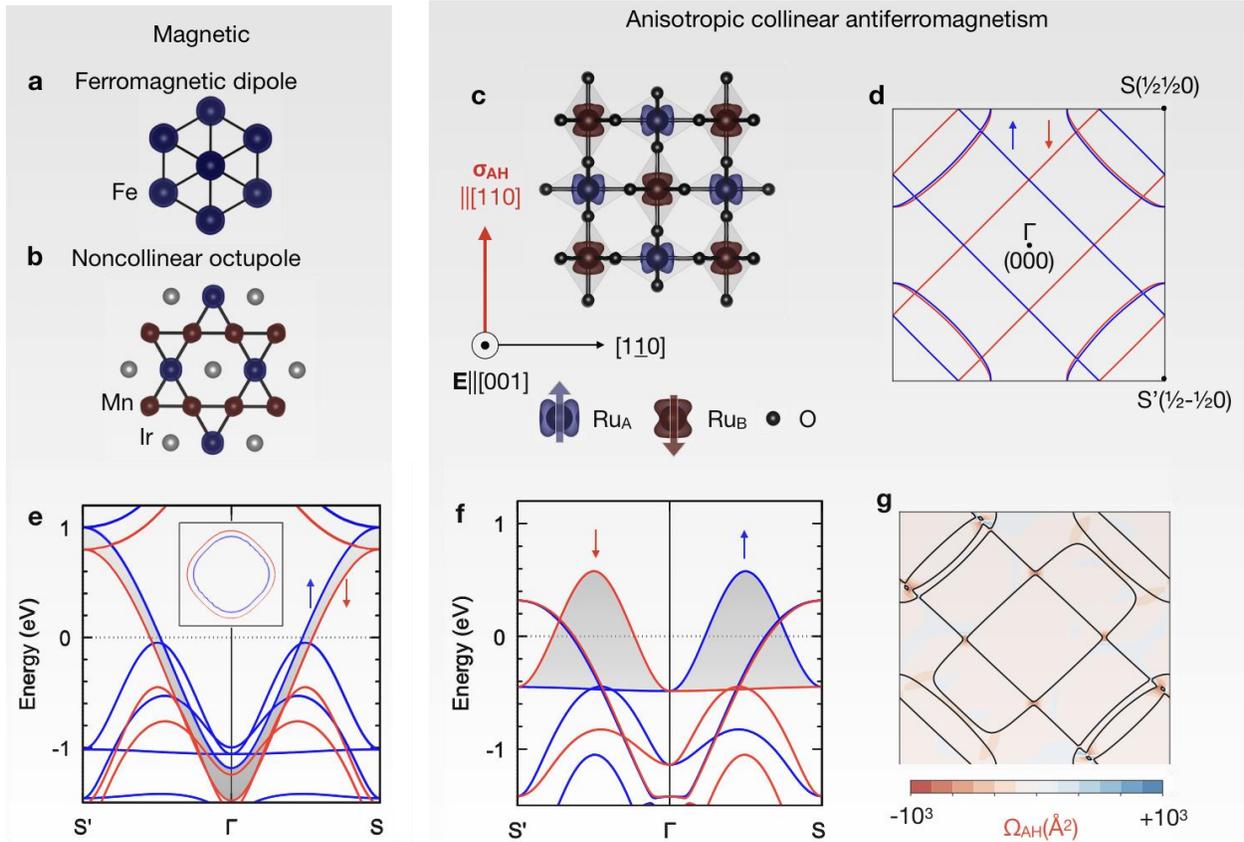

**Figure 1 | Anisotropic collinear antiferromagnetism and anomalous Hall Berry curvature in RuO$_2$ from first principles**. **a & b,** Broadly explored time-reversal symmetry breaking mechanism with ferromagnetic dipole (example of Fe magnetization density) and noncollinear octupole (example of IrMn$_3$ magnetization densities). **c,** The side view of crystal structure of RuO$_2$ with the two Ru antiferromagnetic sublattices and shown O octahedra. Collinear antiferromagnetism with low symmetric magnetization isosurfaces can break time-reversal symmetry. In turn, antiferromagnetic RuO$_2$ with moment along [110] crystal axis exhibits anomalous Hall vector **σ** along [110] and an anomalous Hall voltage can be measured along [1$\bar{1}$0] for electric field **E** applied along [001]. In panels **d-f**, spin-orbit coupling is set to zero. **d,** Calculated spin-split Fermi surfaces shaped as magnetic toroidal quadrupole. **e,** Energy bands of RuO$_2$ in an artificial ferromagnetic state exhibits conventional ferromagnetic spin splitting (gray shading and Fermi surface shown in the inset). **f,** Strong momentum dependent spin splitting (gray shading). **g,** Strong crystal momentum resolved Berry curvature in RuO$_2$ calculated with spin-orbit coupling and moments along [110] corresponding to intrinsic anomalous Hall conductivity vector along [110]. The black contours mark Fermi surface. We set in all first principle calculations $U$ = 1.6 eV and Fermi level to zero.

**Figure 2**

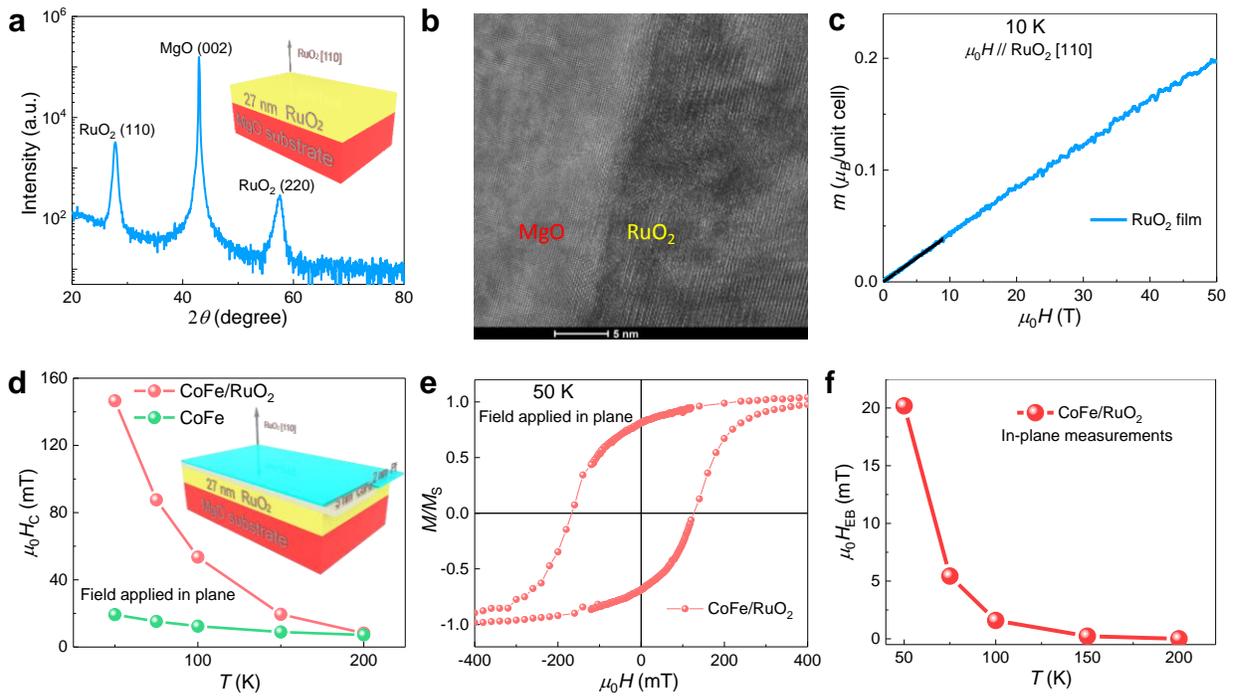

**Figure 2 | Antiferromagnetic order in a RuO$_2$ thin film. a,** X-ray diffraction spectrum of an optimized RuO$_2$/MgO film, indicating a highly ordered (110) orientation of the RuO$_2$ film. Inset: Schematic of the RuO$_2$/MgO heterostructure. **b,** Cross-section transmission electron microscopy image of an optimized RuO$_2$/MgO. **c,** Out-of-plane magnetic moment versus magnetic field of the RuO$_2$/MgO film at 10 K. The black line represents the static-field data measured up to 9 T. **d,** Comparison of the in-plane coercivity field of Co$_{90}$Fe$_{10}$ (CoFe) films in a Pt/CoFe/RuO$_2$/MgO heterostructure and a Pt/CoFe/MgO heterostructure. Inset: Schematic of a RuO$_2$/MgO heterostructure capped by a 5-nm-thick CoFe layer and a 2-nm-thick Pt top layer. **e,** Normalized in-plane magnetization of the Pt/CoFe/RuO$_2$/MgO stack as a function of magnetic field at 50 K. **f,** In-plane exchange bias field of the Pt/CoFe/RuO$_2$/MgO heterostructure versus temperature.

**Figure 3**

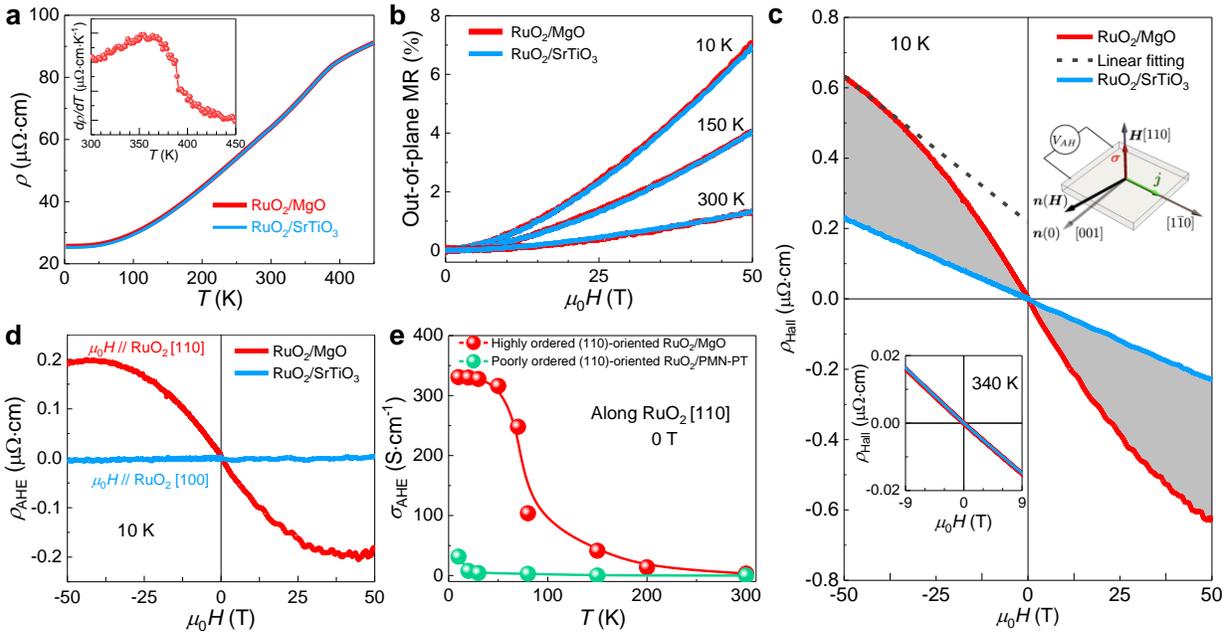

**Figure 3 | Hall effect along the out-of-plane direction of RuO$_2$ thin films. a,** Resistivity versus temperature for (110)-oriented RuO$_2$/MgO and (100)-oriented RuO$_2$/SrTiO$_3$. The residual resistivity ratio defined as $\rho(300K)/\rho(2K)$ is 2.49 and 2.52 for RuO$_2$/MgO and RuO$_2$/SrTiO$_3$, respectively. Inset: Temperature derivative of the resistivity highlighting the Néel temperature transport anomaly. **b,** Out-of-skplane longitudinal magnetoresistance up to 50 T in the two films at different temperatures. **c**, Hall signal of the two films up to 50 T measured at 10 K (for other temperatures see Supplementary Fig. 7 & 8). The grey shaded region corresponds to the nonlinear anomalous Hall resistivity. In the top inset we sketch the experimental setup (**j** and **H** refer to the applied electric current and magnetic field, respectively. **n** and **σ** are the Néel and the Hall vector, respectively. $V_{AH}$ is the measured Hall voltage.). Hall signals measured at 340 K are shown in the bottom inset. **d,** Anomalous Hall resistivity $\rho_{AHE}$ obtained by subtracting from the total Hall signal a linear Hall component corresponding to the linear fit between -40 and -50 T (see Supplementary Fig. 7). **e,** The anomalous Hall conductivity $\sigma_{AHE}$ as a function of temperature for high crystal quality RuO$_2$/MgO and for lower quality RuO$_2$/PMN-PT (see also Supplementary Fig. 9 & 10).

**Figure 4**

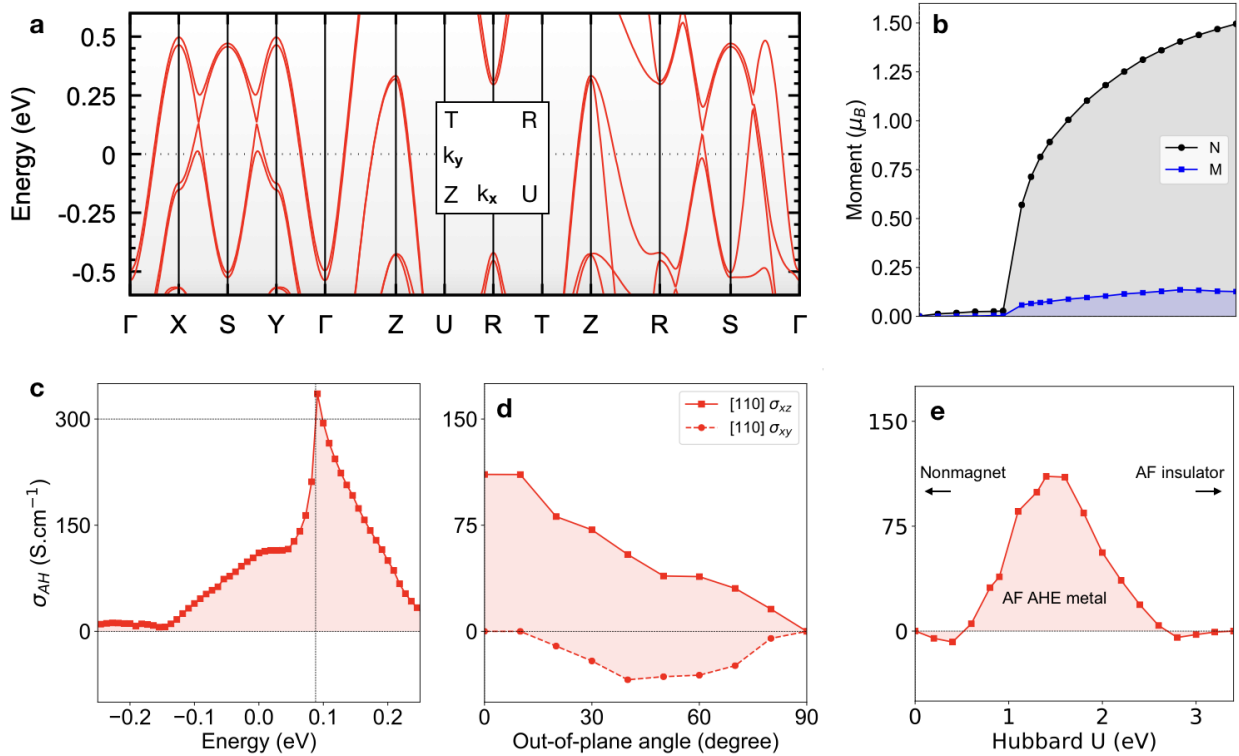

**Figure 4 | Anomalous Hall conductivity of RuO$_2$ calculated from first principles. a,** Energy bands of RuO$_2$ calculated with moments along the [110] direction and spin-orbit interaction (the negligible influence of the magnetic field is shown in Supplementary Fig. 2). In the inset we mark the high symmetry points in the Brillouin zone in the momentum plane $k_z = \pi$. **b,** The magnitude of the antiferromagnetic Néel vector and net moment as a function of the electronic correlation parameter, Hubbard $U$. **c,** Anomalous Hall conductivity versus energy calculated from Wannier Hamiltonian. **d,** Anomalous Hall conductivity components for the antiferromagnetic vector rotation from the [110] to [001] crystal axes. **e,** Anomalous Hall conductivity dependence on the electronic correlation parameter. Where not specified, Hubbard $U$ = 1.6 eV, Néel vector is along the [110] crystal axis and Fermi energy at charge neutrality corresponds to zero.

## Supplementary Figure 1

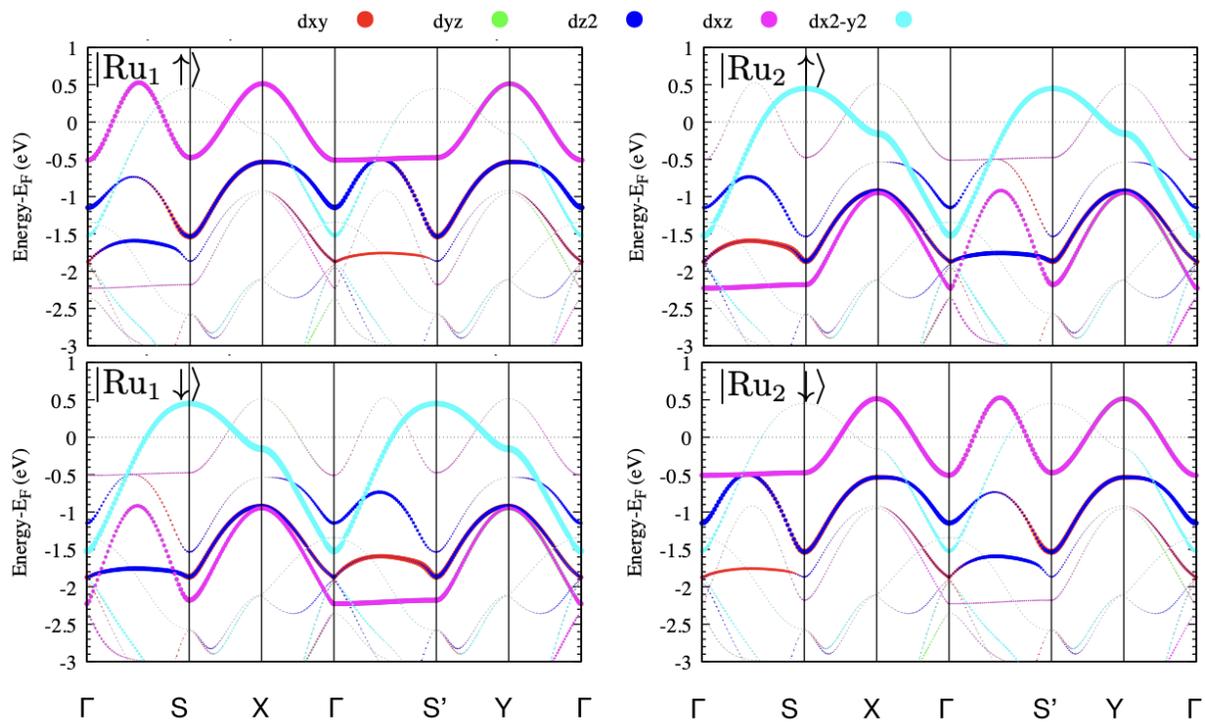

**Supplementary Figure 1 |** Orbital (Ru *d*-states) and spin resolved energy bands in antiferromagnetic state calculated without spin-orbit interaction in VASP.

**Supplementary Figure 2**

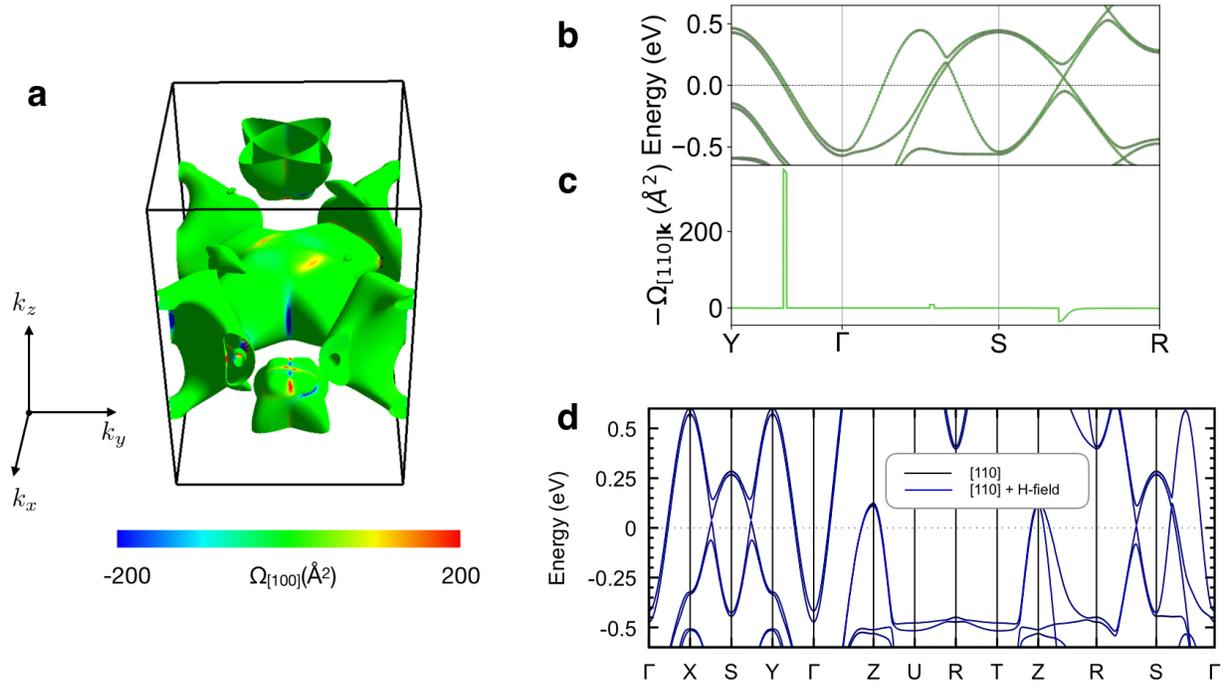

**Supplementary Figure 2 | Detailed Berry curvature calculations and energy bands in Zeeman magnetic field. a,** Berry curvature component -Ω[100] calculated in Wannier90 code[S1] on Fermi surface. **b**, Energy bands and c, Berry curvature -Ω[110] calculated from Wannier Hamiltonian on high symmetry lines in Wannier90 code. All plots use Hubbard $U$ = 1.6 eV and the antiferromagnetic vector along the $RuO_2$[110] crystal axis. Large Berry curvature arises from spin-orbit coupling gapped nodal feature along the ΓX (ΓY) line (blue color in A). These lines are spin degenerate without spin-orbit interaction. Detailed band structures of $RuO_2$ without spin-orbit coupling can be seen in the Supplementary Fig. S6 of our previous theoretical work[S2]. **d**, Energy bands along high symmetry lines calculated for the Néel vector along [110] (black line) and with addition of Zeeman external magnetic field H ~35 T. The panel was calculated in ELK code.

**Supplementary Figure 3**

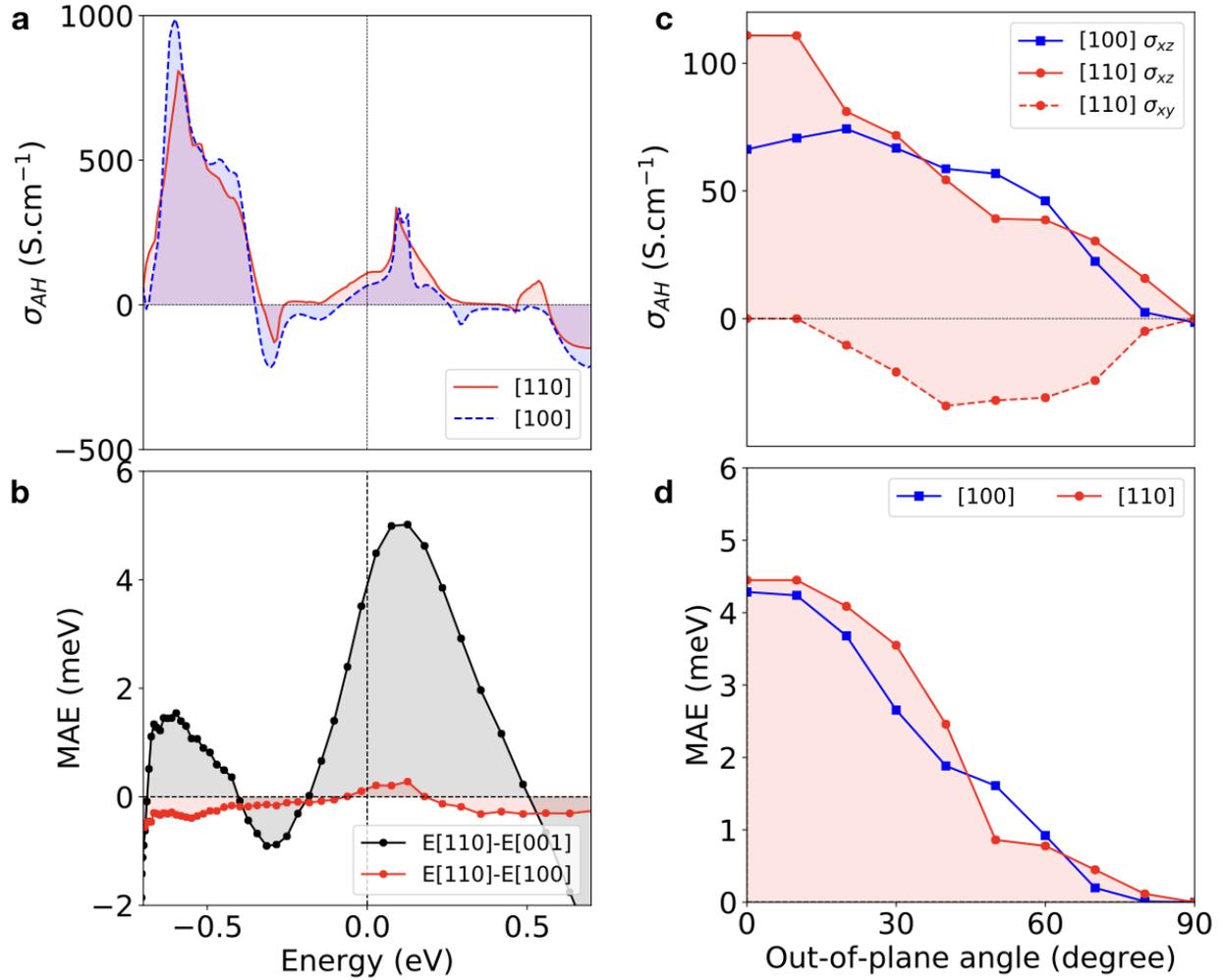

**Supplementary Figure 3 | Detailed anomalous Hall conductivity and magnetic anisotropy energy calculations from first principles. a,** Anomalous Hall conductivity vs energy calculated for the antiferromagnetic vector along [110] and [100] crystal axis in broader energy window. **b,** Corresponding out-of-plane (E[N||110]-E[N||001]) and in-plane (E[N||110]-E[N||100]) magnetic anisotropy energy (MAE) vs Fermi energy. Calculated dependence of the **c,** anomalous Hall conductivity components and **d,** MAE on the antiferromagnetic vector rotation from [100], to [001] and [110] to [001] crystal axis, respectively.

**Supplementary Figure 4**

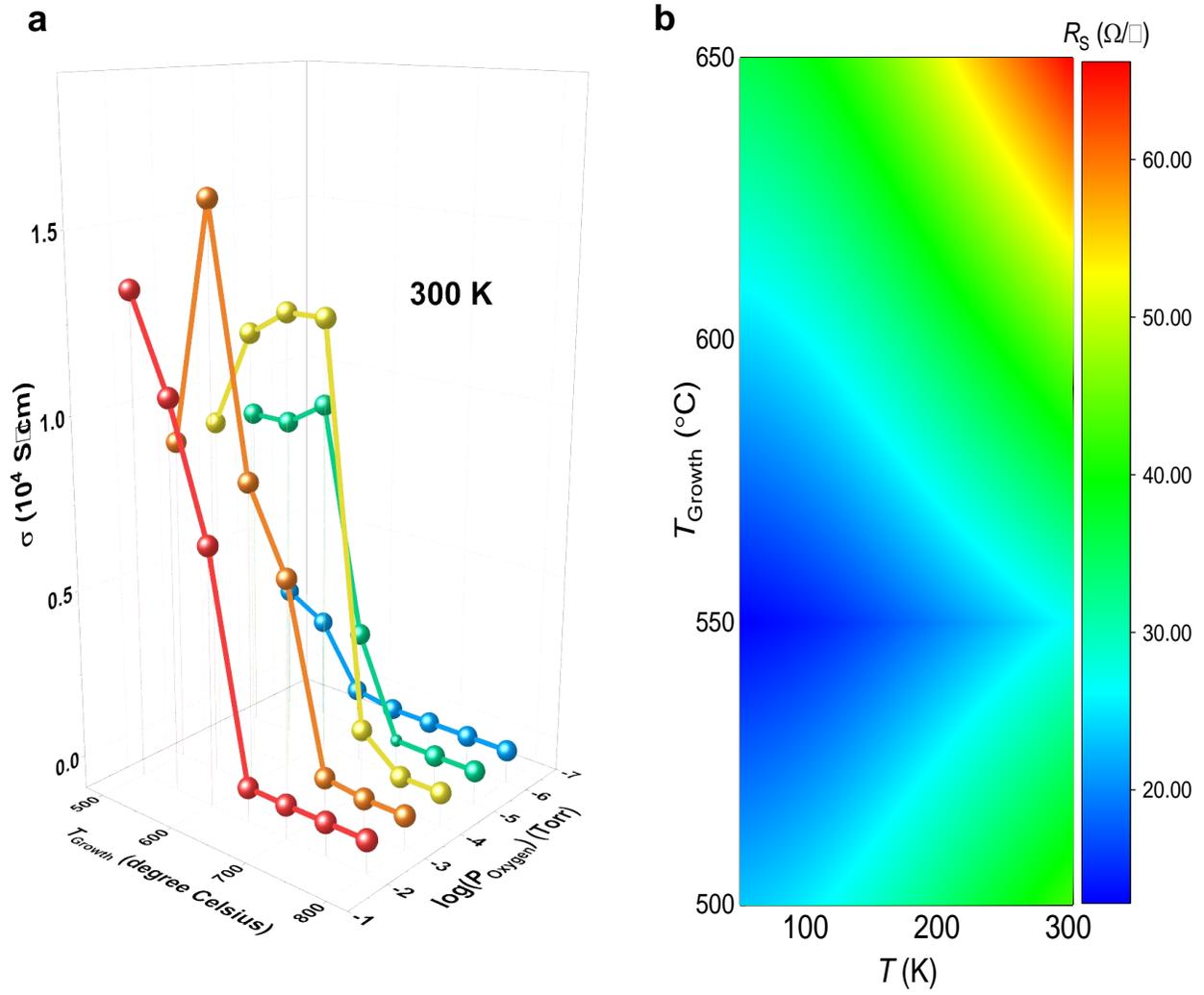

**Supplementary Figure 4 | Conductivity optimization and characterization of RuO$_2$ thin films on MgO (**ref. [S3]**). a.** Phase diagram of electrical conductivity of RuO$_2$ thin films fabricated by pulsed laser deposition at different oxygen pressures and deposition temperatures. **b.** Temperature-dependent sheet resistance contour mapping of the RuO$_2$ films fabricated in the growth temperature range between 500 and 650°C.

**Supplementary Figure 5**

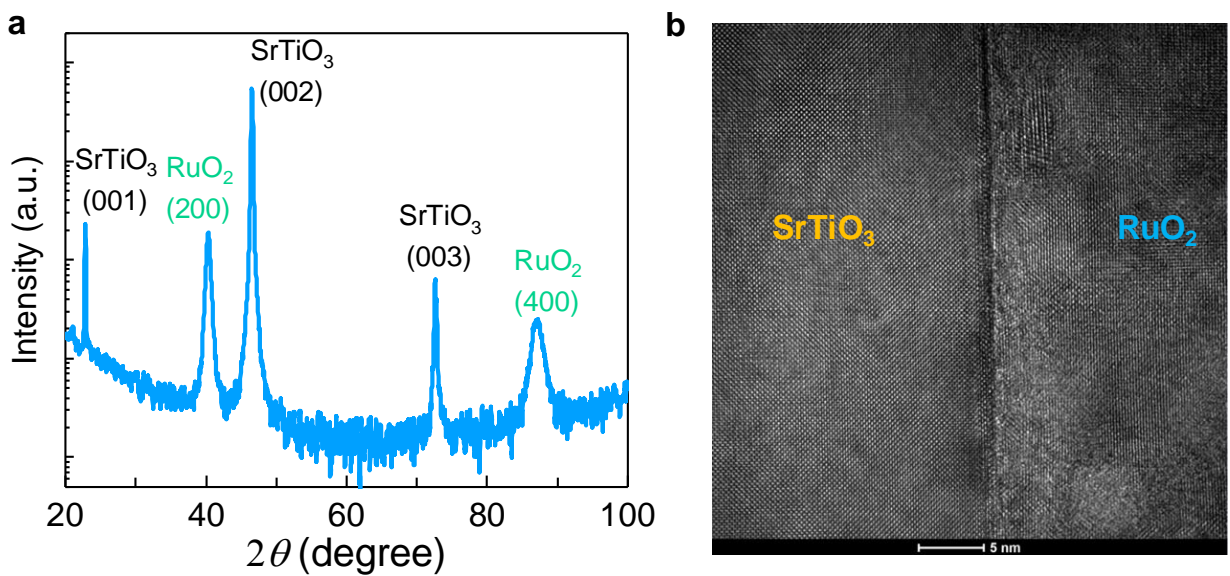

**Supplementary Figure 5 | Structural characterization** of a $RuO_2$/$SrTiO_3$ heterostructure fabricated at 550°C and $10^{-3}$ Torr oxygen partial pressure. **a.** X-ray diffraction pattern. **b.** Cross-section transmission electron microscopy image.

**Supplementary Figure 6**

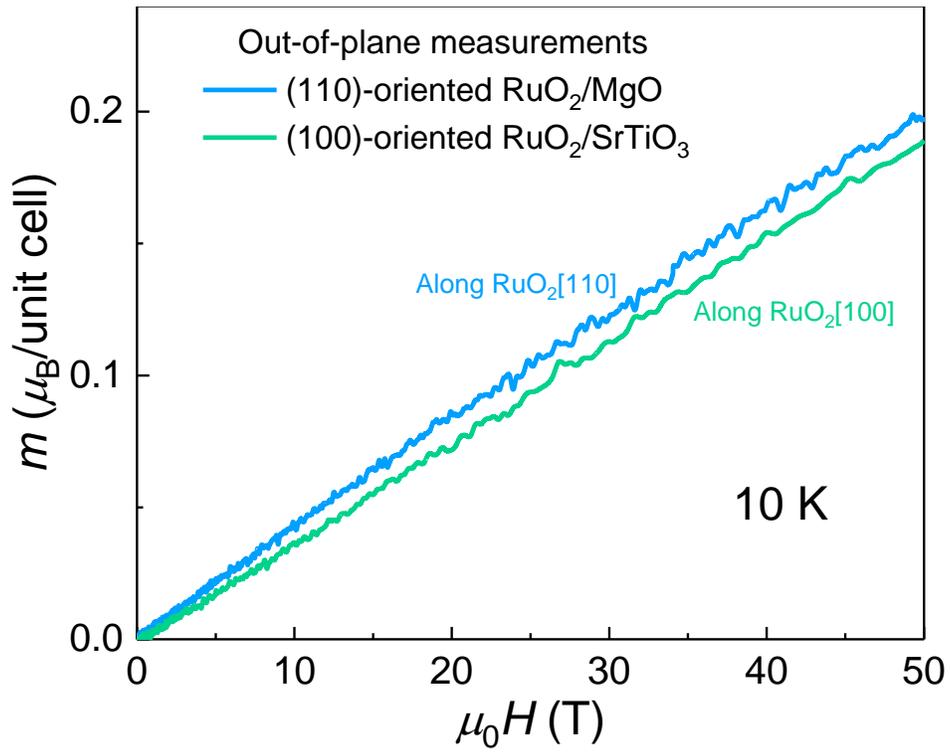

**Supplementary Figure 6 | Out-of-plane magnetic moments** of $RuO_2$ thin films obtained by pulsed-field measurements at 10 K.

**Supplementary Figure 7**

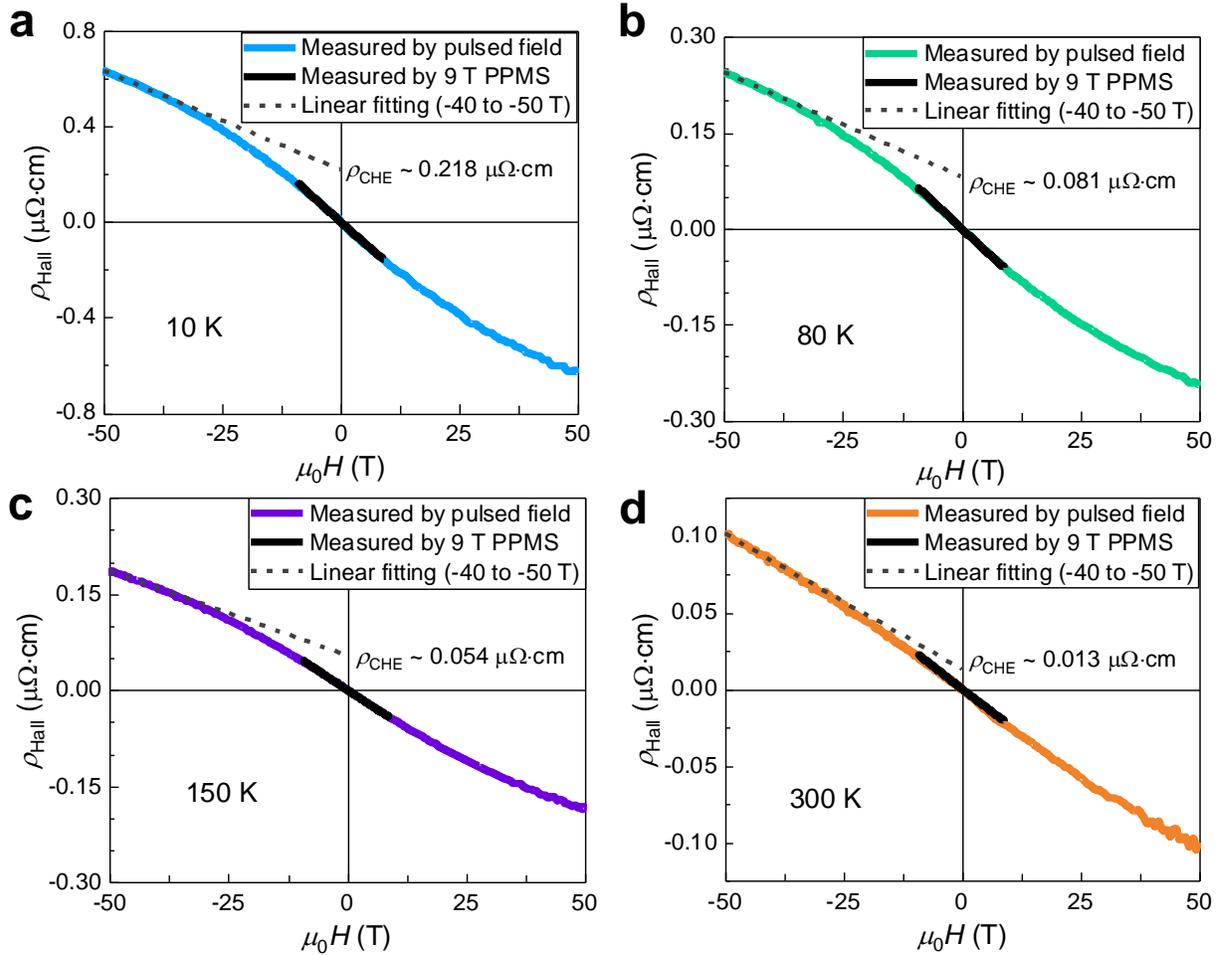

**Supplementary Figure 7 | Total Hall signal** for the $RuO_2$/MgO film obtained by pulsed-field and static-field measurements at different temperatures. Dashed lines represent the linear fitting curves of the total Hall data between -40 and -50 T and the extrapolated zero-field Hall resistivity values are used to represent the zero-field crystal Hall resistivities.

**Supplementary Figure 8**

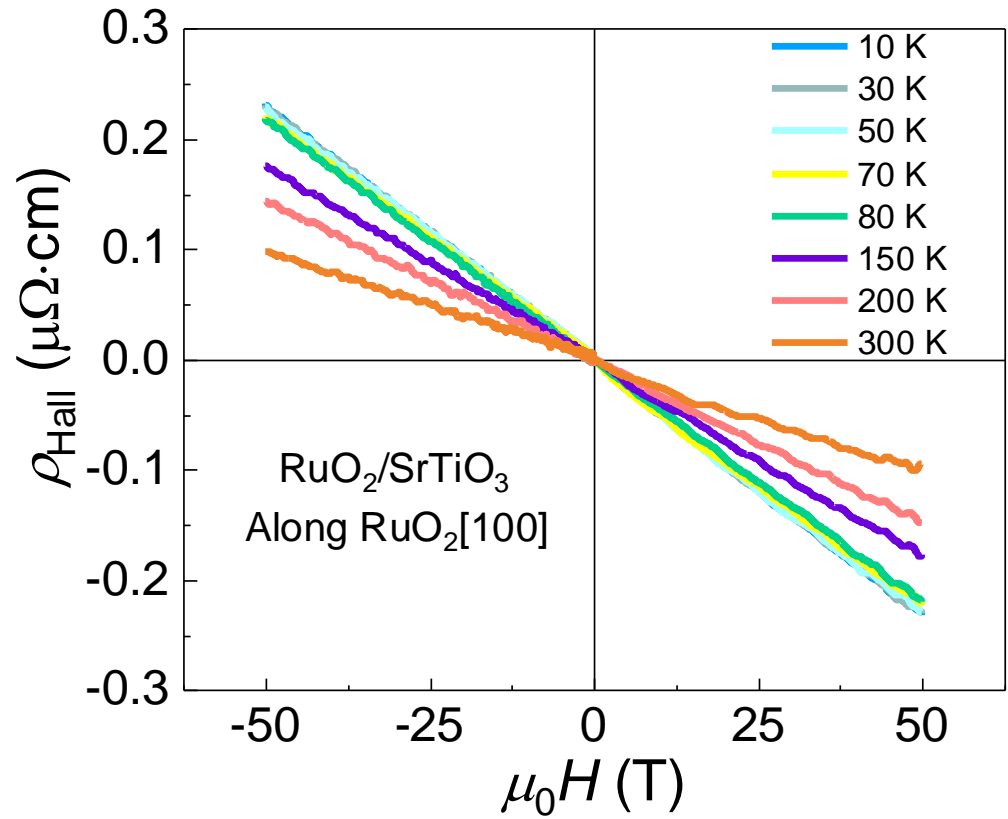

**Supplementary Figure 8 | Total Hall signal** of the (100)-oriented $RuO_2$/$SrTiO_3$ film measured at various temperatures.

## Supplementary Figure 9

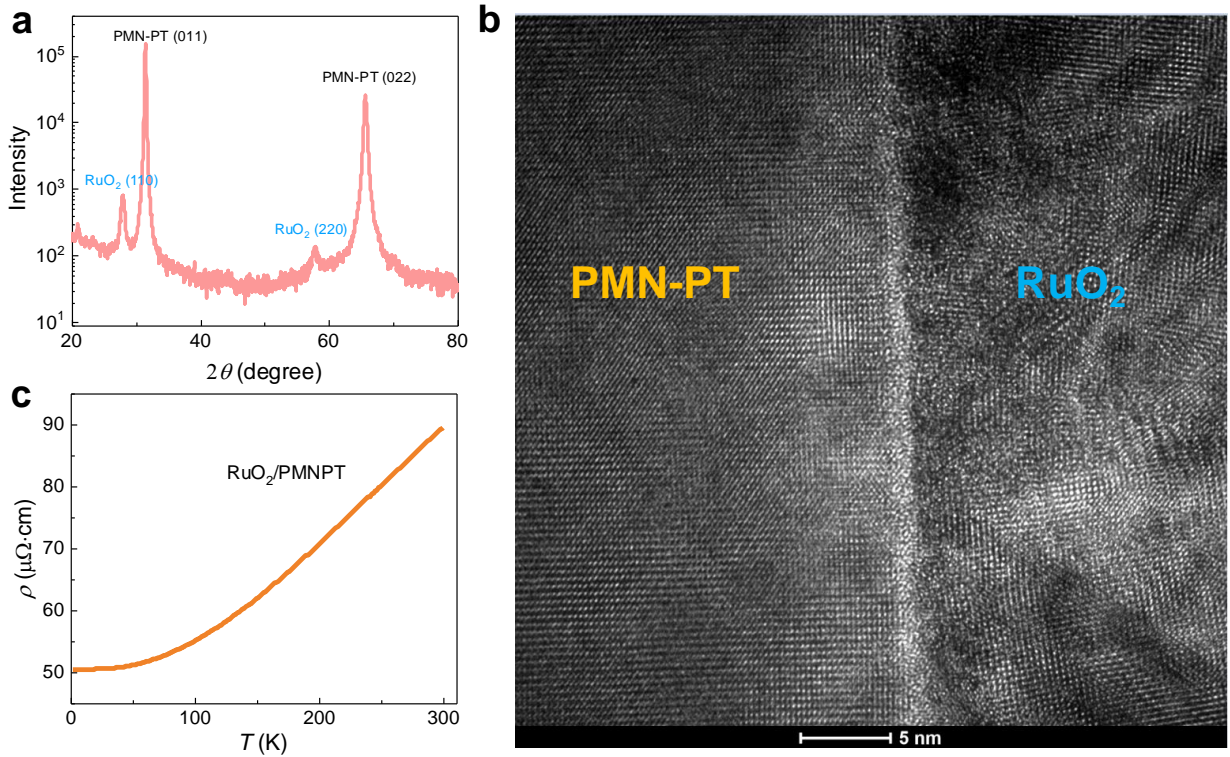

**Supplementary Figure 9 | RuO₂ film grown on a (011)-oriented PMN-PT (0.7PbMg$_{1/3}$Nb$_{2/3}$O$_3$–0.3PbTiO$_3$) single-crystal substrate deposited at 550°C and 10$^{-3}$ Torr oxygen partial pressure. a.** X-ray diffraction spectrum of the RuO$_2$/PMN-PT heterostructure, indicating the (110) orientation of the RuO$_2$ film[S3]. **b.** Cross-section transmission electron microscopy image of the RuO$_2$/PMN-PT heterostructure. **c.** Temperature-dependent resistivity of the RuO$_2$/PMN-PT heterostructure.

**Supplementary Figure 10**

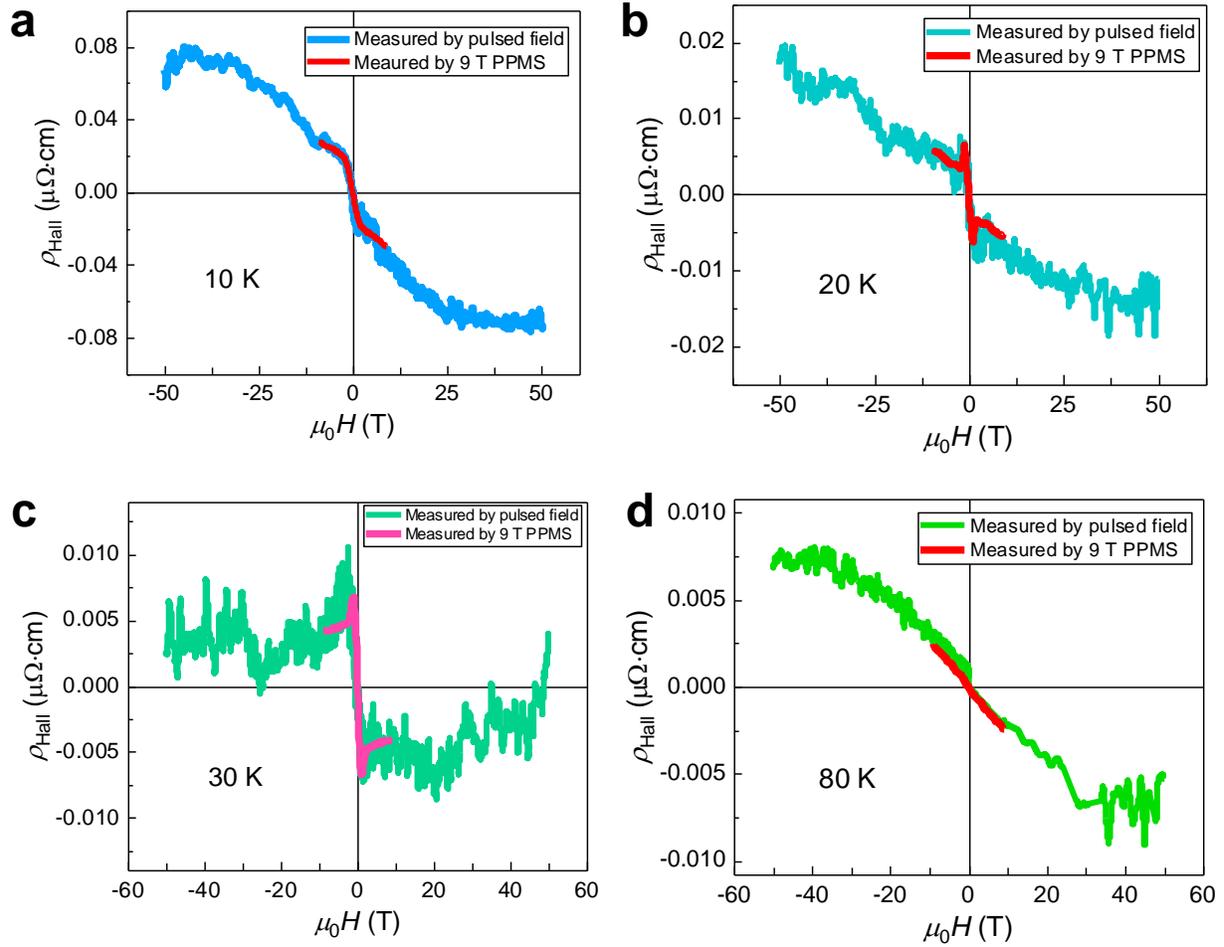

**Supplementary Figure 10 | Crystal Hall effect up to 50 T** along the out-of-plane direction of the (110)-oriented $RuO_2$/PMN-PT heterostructure obtained by pulsed-field and 9 T-PPMS measurements at different temperatures.

**Supplementary Table 1 | Comparison of anomalous Hall effects in antiferromagnets.**

| Hall effect | Antiferromagnet | Longitudinal resistivity $\rho_{xx}$ (μΩ·cm) | Anomalous Hall resistivity $\rho_{AHE}$ (μΩ·cm) | Anomalous Hall conductivity $\sigma_{AHE}$ (S·cm$^{-1}$) |
|---|---|---|---|---|
| Topological | $Mn_5Si_3$ (Ref. [S4]) | 240 | 0.001-0.04 | 0.02-1 |
| Anomalous | $Mn_3Sn$ (Ref. [S5]) | 200 | 4 | 100 |
| Anomalous | $RuO_2$ (Present work) | 26 | 0.21 | 330 |

**Supplementary Table 2 | Physical parameters for the RuO$_2$/MgO heterostructure.**

| Temperature (K) | $\rho_{xx}$ under zero field ($\mu\Omega\cdot$cm) | $\rho_{AHE}$ extrapolated to zero field ($\mu\Omega\cdot$cm) | $\sigma_{AHE}$ for zero field (S·cm$^{-1}$) | Single-band carrier density estimation based on the -40~-50 T linear slope (holes/cm$^3$) | Single-band mobility estimation based on the -40~-50 T linear slope (cm$^2\cdot$V$^{-1}\cdot$s$^{-1}$) |
|---|---|---|---|---|---|
| 10 | 25.67 | 0.218 | 330.83 | 0.75×10$^{23}$ | 3.24 |
| 80 | 27.94 | 0.081 | 103.76 | 1.92×10$^{23}$ | 1.16 |
| 150 | 36.19 | 0.054 | 41.23 | 2.35×10$^{23}$ | 0.73 |
| 300 | 63.74 | 0.013 | 3.20 | 3.53×10$^{23}$ | 0.28 |

## Supplementary References